\documentclass[conference]{IEEEtran}

\usepackage[utf8]{inputenc}
\usepackage[T1]{fontenc}
\usepackage{amsmath,amssymb}
\usepackage{graphicx}
\usepackage{tikz}
\usetikzlibrary{positioning, arrows.meta, shapes.geometric, fit, calc, decorations.pathreplacing}
\usepackage{listings}
\usepackage{xcolor}
\usepackage{url}

\usepackage{hyperref}
\usepackage{booktabs}
\usepackage{pifont}
\usepackage{algorithm}
\usepackage{algorithmic}
\usepackage{balance}
\usepackage{pgfplots}
\pgfplotsset{compat=1.18}

\hypersetup{colorlinks=true, linkcolor=blue, citecolor=blue, urlcolor=blue}

\title{A Three-Layer Caching Architecture for Low-Latency LLM Web Search on Commodity CPU Hardware}

\author{
\IEEEauthorblockN{Ayushman Bhattacharya}
\IEEEauthorblockA{pollinations.ai\\ayushman@pollinations.ai}
\and
\IEEEauthorblockN{Nihal Gazi}
\IEEEauthorblockA{pollinations.ai\\info@nihalgazi.com}
}

\begin{document}

\maketitle

\begin{abstract}
AI-powered search products such as ChatGPT search, Google's AI Overviews, and Perplexity provide LLM-synthesized answers grounded in live web results. We developed OreoLook (formerly lixSearch), an open-source answer engine using automated browser agents and provider-routed LLM inference. Its local search, caching, session-management, and embedding stack runs on commodity CPU hardware; answer synthesis is performed by a remote inference provider. As usage grew, sessions lost context, equivalent queries triggered redundant work, and URLs were repeatedly embedded across sessions.

We present a three-layer caching architecture: (1)~a \emph{Session Context Window} maintaining a rolling window of recent messages in Redis with automatic overflow to Huffman-compressed disk archives; (2)~a \emph{Semantic Query Cache} catches rephrasings via cosine similarity on embedding vectors, eliminating redundant LLM invocations; and (3)~a \emph{URL Embedding Cache} that deduplicates embedding computations across sessions. Deployed on a single 8-vCPU Intel Cascade Lake server (2\,GHz, 32\,GB RAM) running 30 Hypercorn worker processes across three containerized replicas, the evaluated system reported an 89.3\% aggregate Redis keyspace hit rate with 0.1\,ms read latency and just 1.38\,MB of memory overhead. A background LRU eviction daemon migrates idle sessions from Redis to disk and re-hydrates them on demand, enabling conversations that can be resumed hours or days later under the configured retention policy.
\end{abstract}


\section{Introduction}

\subsection{The Problem: AI Search APIs Are Too Expensive}

The past two years have brought a wave of AI-powered search products. OpenAI launched SearchGPT in late 2024 (now integrated into ChatGPT). Google added AI Overviews with Gemini grounding. Perplexity created a real-time answer engine with its Sonar API.

The API pricing across providers is very sharp. OpenAI charges \$10--30 per thousand search calls as a base fee (varying by model and context tier), plus ${\sim}$8{,}000 tokens of injected web context billed at standard input rates. In practice, a single query costs \$0.03--0.10 depending on the model used~\cite{openai_search_pricing}. Developers on OpenAI's community forums reported bills 2--3$\times$ higher than expected, with some paying over \$2 for just 21 searches (${\sim}$\$0.10 each)~\cite{openai_search_billing}. Google's Gemini with grounding charges \$35 per 1{,}000 grounded queries for the Pro model (\$14/1K for Flash)~\cite{google_grounding_pricing}. Perplexity's Sonar API charges \$5 per 1{,}000 requests for the base search tier, scaling to \$18/1K for Sonar Pro~\cite{perplexity_pricing}.

We wanted to build something different: a search assistant that could browse the web, synthesize answers with sources, and carry on multi-turn conversations, but at a cost we could actually sustain.

\subsection{The First Version: Raw Search + LLM}

So we built OreoLook (formerly lixSearch) from scratch. The initial version was deliberately simple. Instead of paying per-query fees to a proprietary search API, we used automated headless browser agents to perform web searches directly---the same way a human would. The agents navigated search engines, extracted results, fetched full-page content, and sent it to provider-routed LLM inference for synthesis. There was no retrieval-augmented generation (RAG), no vector database, no caching layer. Just a search agent pool, a language model, and a pipeline connecting them.

It worked. The per-query cost dropped dramatically---from \$0.03--0.10 with SearchGPT to ${\sim}$\$0.02 with open-provider LLM inference (without per-query search API fees; only provider inference was billed). But as users grew and conversations got longer, three problems emerged that threatened the cost advantage we had built:

\begin{enumerate}
\item \textbf{``What did we just talk about?''} Users expected the system to remember context across turns. Storing entire conversation histories in memory did not scale across thousands of concurrent sessions. Without context, the assistant repeated itself, missed follow-up nuances, and frustrated users.

\item \textbf{``Didn't we already answer this?''} Users frequently rephrased queries: ``weather Tokyo'' followed by ``Tokyo weather forecast.'' Each rephrasing triggered a full pipeline execution---search agents, page fetches, LLM synthesis---even though the answer was already computed seconds ago.

\item \textbf{``We already embedded this URL.''} As we added RAG capabilities to improve answer quality, the same popular URLs were fetched and embedded by multiple sessions. Computing a 384-dimensional embedding (${\sim}200$\,ms per URL) redundantly across sessions wasted the compute budget we had fought so hard to minimize.
\end{enumerate}

\subsection{System and Artifact Naming}

The deployed answer engine is named \emph{OreoLook}; earlier versions and historical measurements used \emph{lixSearch}. The reusable cache implementation evaluated here is \texttt{lix-open-cache}. We do not bind the design to a transient synthesis-model alias: provider models are selected through a routing layer, while the evaluated local embedding model is \texttt{sentence-transformers/all-MiniLM-L6-v2}.

\subsection{The Solution: A Three-Tier Cache Architecture}

Each of these problems had partial solutions in the ecosystem. LangChain~\cite{langchain2023} offered in-memory conversation buffers. GPTCache~\cite{gptcache2023} provided semantic caching for LLM responses. But nothing unified all three concerns into a single, lightweight system we could integrate into our existing pipeline without adding heavyweight infrastructure.

So we built a three-layer caching architecture that grew organically out of the problems we faced in production. Each layer was created due to a specific pain point:

\begin{itemize}
\item \textbf{Layer 1: Session Context Window.} Rolling window in Redis with Huffman-compressed disk overflow.
\item \textbf{Layer 2: Semantic Query Cache.} Catches rephrasings via cosine similarity, skipping the entire pipeline on a hit.
\item \textbf{Layer 3: URL Embedding Cache.} Global cross-session store of pre-computed embedding vectors.
\end{itemize}

\begin{figure}[h]
\centering
\resizebox{\columnwidth}{!}{%
\begin{tikzpicture}[
    node distance=0.5cm,
    box/.style={draw, rounded corners=3pt, minimum height=0.6cm, minimum width=2.8cm, font=\scriptsize, align=center},
    cache/.style={box, fill=green!12, draw=green!50},
    pipe/.style={box, fill=blue!8, draw=blue!40},
    hit/.style={box, fill=yellow!20, draw=yellow!60},
    arr/.style={-{Stealth[length=4pt]}, thick},
]
\node[pipe] (query) {User Query};
\node[cache, below=0.5cm of query] (l2) {Layer 2: Semantic Cache\\cosine sim $\geq 0.90$?};
\node[hit, right=1.5cm of l2] (cached) {Return cached\\response};
\node[pipe, below=0.5cm of l2] (ctx) {Layer 1: Load session\\context ($k{=}20$)};
\node[cache, below=0.5cm of ctx] (l3) {Layer 3: URL Embedding\\Cache (24\,h TTL)};
\node[pipe, below=0.5cm of l3] (tools) {Search agents +\\page fetches};
\node[pipe, below=0.5cm of tools] (llm) {LLM Synthesis};
\node[pipe, below=0.5cm of llm] (resp) {Response + Sources};

\draw[arr] (query) -- (l2);
\draw[arr, green!60] (l2) -- (cached) node[midway, above, font=\tiny] {HIT};
\draw[arr] (l2) -- (ctx) node[midway, right, font=\tiny] {MISS};
\draw[arr] (ctx) -- (l3);
\draw[arr] (l3) -- (tools);
\draw[arr] (tools) -- (llm);
\draw[arr] (llm) -- (resp);
\end{tikzpicture}%
}
\caption{Query pipeline flow. A semantic cache hit (Layer~2) short-circuits the entire pipeline. On a miss, session context is loaded (Layer~1), embeddings are checked (Layer~3), and the full search--synthesis pipeline executes.}
\label{fig:pipeline}
\end{figure}
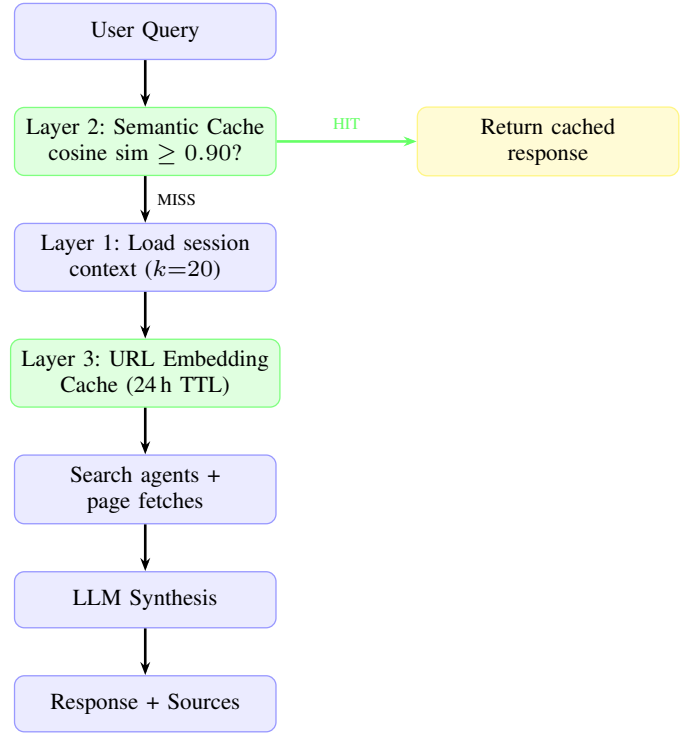

Fig.~\ref{fig:pipeline} illustrates how a user query flows through these layers before reaching the LLM.

The remainder of this paper follows the setup of this journey. Section~\ref{sec:related} surveys related work, Section~\ref{sec:architecture} presents the architecture, Section~\ref{sec:design} explains the key design decisions, Section~\ref{sec:implementation} details the implementation, Section~\ref{sec:evaluation} evaluates production performance, and Section~\ref{sec:conclusion} concludes with limitations and future directions.

\section{Related Work}
\label{sec:related}

Table~\ref{tab:related} summarizes the landscape of existing tools and how our system differs. No single prior system addresses all three concerns (session persistence, semantic deduplication, embedding reuse) in a unified, lightweight package.

\begin{table}[h]
\centering
\caption{Feature comparison with existing systems}
\label{tab:related}
{\scriptsize
\begin{tabular}{@{}lccccc@{}}
\toprule
& \rotatebox{60}{\textbf{Session persist.}} & \rotatebox{60}{\textbf{Disk archival}} & \rotatebox{60}{\textbf{Semantic dedup}} & \rotatebox{60}{\textbf{Embed. reuse}} & \rotatebox{60}{\textbf{Redis-only}} \\
\midrule
LangChain~\cite{langchain2023} & \ding{55} & \ding{55} & \ding{55} & \ding{55} & \ding{55} \\
LlamaIndex~\cite{llamaindex2023} & \ding{55} & \ding{55} & \ding{55} & \ding{55} & \ding{55} \\
GPTCache~\cite{gptcache2023} & \ding{55} & \ding{55} & \ding{51} & \ding{55} & \ding{55} \\
MemGPT~\cite{memgpt2023} & \ding{51} & \ding{51} & \ding{55} & \ding{55} & \ding{55} \\
Sem.\ Kernel~\cite{semantic_kernel2023} & \ding{55} & \ding{55} & \ding{55} & \ding{55} & \ding{51} \\
\midrule
\textbf{Ours} & \ding{51} & \ding{51} & \ding{51} & \ding{51} & \ding{51} \\
\bottomrule
\end{tabular}}
\end{table}

\textbf{Conversation memory in LLM frameworks.}
LangChain~\cite{langchain2023} provides several memory modules---buffer, summary, and windowed variants---that maintain conversation state for LLM chains. These operate in-process and do not persist across restarts or scale across replicas. LlamaIndex~\cite{llamaindex2023} offers similar in-memory chat stores. Our system differs by providing Redis-backed persistence with automatic disk archival, enabling shared state across horizontally scaled application instances.

\textbf{Semantic caching for LLMs.}
GPTCache~\cite{gptcache2023} is the closest prior work to our semantic query cache layer. It intercepts LLM calls, computes embeddings for queries, and returns cached responses when similarity exceeds a threshold. GPTCache supports multiple embedding backends and vector stores (FAISS, Milvus, etc.). Our design differs in three significant ways: (1)~GPTCache is a \emph{global} cache with no per-session isolation---in a multi-user search assistant, this would leak responses across users. Our cache is scoped per-session by design. (2)~GPTCache requires a separate vector database (FAISS, Milvus, or Qdrant) for similarity search, adding operational complexity. Our system stores embeddings directly in Redis as JSON arrays, requiring no additional infrastructure. (3)~GPTCache addresses only the semantic caching concern. It does not provide session context management, disk archival, or cross-session embedding reuse. These are the other two layers that, in our experience, account for the majority of compute savings.

We did not benchmark GPTCache directly against our system because the two are not drop-in replacements: GPTCache is a middleware that wraps LLM calls, while our system is an integrated caching layer that manages sessions, context, and embeddings as a unified concern. A meaningful comparison would require building equivalent session management and embedding reuse on top of GPTCache, which would effectively recreate our architecture.

\textbf{Redis as a caching layer.}
Redis is widely used as an LLM response cache. Frameworks like Semantic Kernel~\cite{semantic_kernel2023} and Haystack~\cite{haystack2023} support Redis as a cache backend. However, these typically use Redis as a flat key-value store. Our system uses three separate Redis logical databases with distinct TTL profiles and data formats, and adds the hybrid hot/cold tier with disk overflow---a pattern not found in existing frameworks.

\textbf{Conversation compression and archival.}
MemGPT~\cite{memgpt2023} addresses the context window limitation by paging conversation history between a main context and an external storage tier, analogous to virtual memory. Our approach is similar in spirit: the hot Redis window serves as ``main memory'' and the Huffman-compressed disk archive as ``swap.'' However, MemGPT focuses on autonomous LLM-driven memory management (the LLM decides what to page in/out), while our system uses deterministic LRU eviction with fixed window sizes---simpler to reason about and debug in production.

\textbf{Data compression for chat.}
Standard approaches use gzip or lz4 for compressing stored conversations. Our use of canonical Huffman coding is motivated by the small payload sizes typical of conversation archives ($<$10\,KB in 90\% of cases), where dictionary-based compressors have proportionally higher overhead. As shown in Table~\ref{tab:compression_comparison}, Huffman outperforms lz4 and approaches zlib within 5--10 percentage points at these sizes, while requiring zero native dependencies---simplifying deployment in containerized environments.

\section{Architecture}
\label{sec:architecture}

\subsection{Overview}

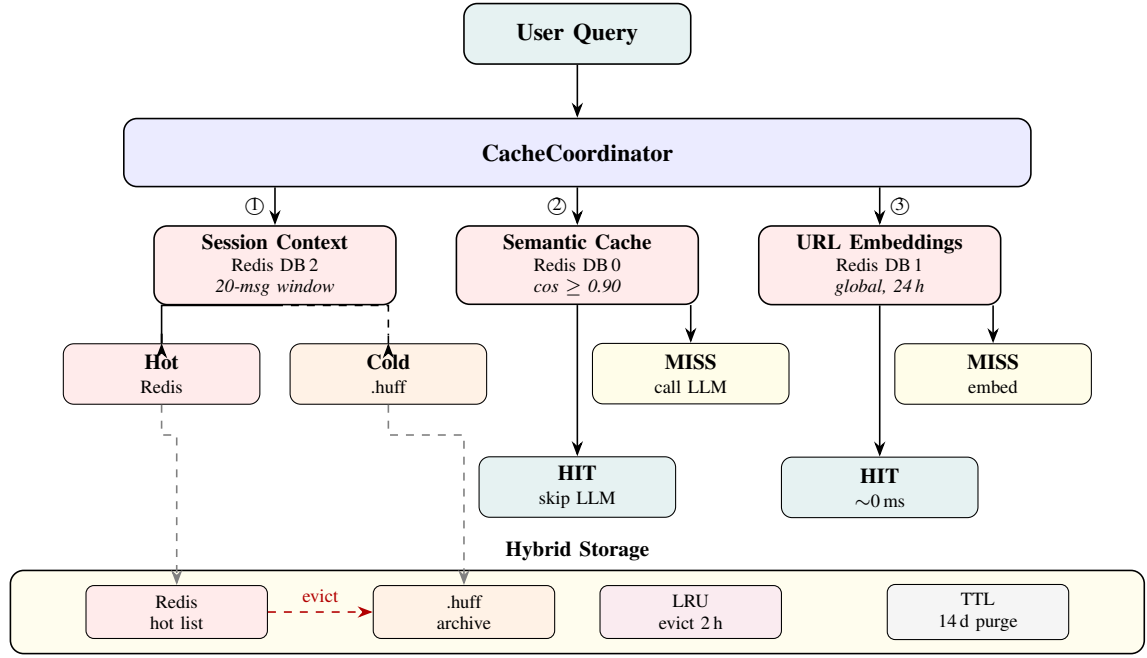
\begin{figure*}[t]
\centering
\begin{tikzpicture}[
    scale=1.0, transform shape,
    >=Stealth,
    every node/.style={font=\small},
    mainbox/.style={draw, rounded corners=5pt, minimum width=3.2cm, minimum height=1.0cm, align=center, semithick},
    subbox/.style={draw, rounded corners=4pt, minimum width=2.6cm, minimum height=0.8cm, align=center, font=\footnotesize, thin},
    smallbox/.style={draw, rounded corners=3pt, minimum width=2.4cm, minimum height=0.7cm, align=center, font=\scriptsize, thin},
    redis/.style={fill=red!8},
    disk/.style={fill=orange!10},
    coord/.style={fill=blue!8},
    tealfill/.style={fill=teal!10},
    yellowfill/.style={fill=yellow!10},
    purplefill/.style={fill=purple!8},
    grayfill/.style={fill=gray!8},
    arrow/.style={->, semithick, >=Stealth},
    dasharrow/.style={->, semithick, dashed, >=Stealth},
]

\node[draw, rounded corners=5pt, fill=teal!10,
      minimum width=3.0cm, minimum height=0.8cm, semithick]
    (user) {\textbf{User Query}};

\node[draw, rounded corners=5pt, coord,
      minimum width=12.0cm, minimum height=0.9cm, semithick,
      below=0.7cm of user]
    (coordinator) {\textbf{CacheCoordinator}};

\node[mainbox, redis,
      below=0.5cm of coordinator, xshift=-4.0cm]
    (session) {%
        \textbf{\footnotesize Session Context}\\[-2pt]
        {\scriptsize Redis DB\,2}\\[-2pt]
        {\scriptsize\textit{20-msg window}}};

\node[mainbox, redis,
      below=0.5cm of coordinator]
    (semantic) {%
        \textbf{\footnotesize Semantic Cache}\\[-2pt]
        {\scriptsize Redis DB\,0}\\[-2pt]
        {\scriptsize\textit{cos $\geq$ 0.90}}};

\node[mainbox, redis,
      below=0.5cm of coordinator, xshift=4.0cm]
    (urlcache) {%
        \textbf{\footnotesize URL Embeddings}\\[-2pt]
        {\scriptsize Redis DB\,1}\\[-2pt]
        {\scriptsize\textit{global, 24\,h}}};

\node[subbox, redis,
      below=0.5cm of session, xshift=-1.5cm]
    (hot) {\textbf{Hot}\\{\scriptsize Redis}};

\node[subbox, disk,
      below=0.5cm of session, xshift=1.5cm]
    (cold) {\textbf{Cold}\\{\scriptsize .huff}};

\node[subbox, tealfill,
      below=2cm of semantic, xshift=0cm]
    (semhit) {\textbf{HIT}\\{\scriptsize skip LLM}};

\node[subbox, yellowfill,
      below=0.5cm of semantic, xshift=1.5cm]
    (semmiss) {\textbf{MISS}\\{\scriptsize call LLM}};

\node[subbox, tealfill,
      below=2cm of urlcache, xshift=0cm]
    (urlhit) {\textbf{HIT}\\{\scriptsize $\sim$0\,ms}};

\node[subbox, yellowfill,
      below=0.5cm of urlcache, xshift=1.5cm]
    (urlmiss) {\textbf{MISS}\\{\scriptsize embed}};

\node[draw, rounded corners=5pt, fill=yellow!8,
      minimum width=15.0cm, minimum height=1.1cm, semithick,
      below=3.5cm of semantic,
      label={[font=\footnotesize\bfseries]above:Hybrid Storage}]
    (hybrid) {};

\node[smallbox, redis, at=(hybrid.center), xshift=-5.3cm]
    (hybhot) {Redis\\hot list};

\node[smallbox, disk, at=(hybrid.center), xshift=-1.5cm]
    (hybcold) {.huff\\archive};

\node[smallbox, purplefill, at=(hybrid.center), xshift=1.5cm]
    (hyblru) {LRU\\evict 2\,h};

\node[smallbox, grayfill, at=(hybrid.center), xshift=5.3cm]
    (hybttl) {TTL\\14\,d purge};

\draw[arrow] (user.south) -- (coordinator.north);

\draw[arrow]
    (coordinator.south -| session.north) -- (session.north)
    node[midway, left, font=\scriptsize] {\textcircled{\scriptsize 1}};

\draw[arrow]
    (coordinator.south) -- (semantic.north)
    node[midway, left, font=\scriptsize] {\textcircled{\scriptsize 2}};

\draw[arrow]
    (coordinator.south -| urlcache.north) -- (urlcache.north)
    node[midway, right, font=\scriptsize] {\textcircled{\scriptsize 3}};

\draw[arrow] (session.south) -- ++(-1.5cm, 0) |- (hot.north);
\draw[dasharrow] (session.south) -- ++(1.5cm, 0) |- (cold.north);

\draw[arrow] (semantic.south) -- ++(0, -0.4cm) -| (semhit.north);
\draw[arrow] (semantic.south -| semmiss.north) -- (semmiss.north);

\draw[arrow] (urlcache.south) -- ++(0cm, -0.4cm) -| (urlhit.north);
\draw[arrow] (urlcache.south -| urlmiss.north) -- (urlmiss.north);

\draw[dasharrow, gray]
    (hot.south)  -- ++(0,-0.4cm) -| (hybhot.north);
\draw[dasharrow, gray]
    (cold.south) -- ++(0,-0.4cm) -| (hybcold.north);

\draw[dasharrow, red!70!black]
    (hybhot.east) -- (hybcold.west)
    node[midway, above, font=\scriptsize, red!70!black] {evict};

\end{tikzpicture}
\caption{Data flow through the three-layer caching architecture. Solid arrows indicate the primary path; dashed arrows indicate overflow and eviction paths.}
\label{fig:architecture}
\end{figure*}

With the problems identified, the architecture took shape around a simple 
principle: each caching concern gets its own logical partition within a single Redis~\cite{redis2023} instance (DB~0 for semantic query cache, DB~1 for URL embeddings, DB~2 for session context), and a single coordinator process ties them together. Fig.~\ref{fig:architecture} illustrates 
the complete data flow when a user message arrives.

\subsection{Layer 1: Session Context Window (Redis \texorpdfstring{DB\,2}{DB 2})}

Users expected the assistant to remember what they had said two turns ago. The Session Context Window maintains a rolling window of the $k$ most recent messages (default $k{=}20$) for each session. Messages are stored as individual Redis keys with TTL, and an ordered list tracks message insertion order.

When a new message arrives, it is pushed to the head of the Redis list. If the list exceeds $k$ entries, the oldest message is popped, serialized, and appended to a Huffman-compressed disk archive. This ensures Redis memory usage remains bounded at $O(k)$ per session regardless of conversation length.

When the user query requests context and Redis is empty (e.g., after LRU eviction), the system transparently re-hydrates by loading the last $k$ messages from the disk archive back into Redis. If Redis is entirely unavailable, the system falls back to disk-only reads to ensure no downtime.

\subsection{Layer 2: Semantic Query Cache (Redis \texorpdfstring{DB\,0}{DB 0})}

Users do not type the same query twice-they rephrase it. ``What's the weather in India'' becomes ``India weather forecast'' becomes ``India temperature today.'' Before we established this architecture, each variation triggered a full pipeline run: search agents launched, pages fetched, LLM invoked. The Semantic Query Cache intercepts queries before they reach the LLM. For each incoming query, the system:

\begin{enumerate}
\item Computes an embedding vector $\mathbf{q} \in \mathbb{R}^{384}$ for the query.
\item Retrieves all cached $(\mathbf{e}_i, r_i)$ pairs for the current session and URL, where $\mathbf{e}_i$ is a cached embedding and $r_i$ the corresponding LLM response.
\item Computes cosine similarity: $\text{sim}(\mathbf{q}, \mathbf{e}_i) = \frac{\mathbf{q} \cdot \mathbf{e}_i}{\|\mathbf{q}\| \|\mathbf{e}_i\|}$.
\item If $\max_i \text{sim}(\mathbf{q}, \mathbf{e}_i) \geq \tau$ (default $\tau{=}0.90$), returns the cached response $r_i$ and skips the LLM entirely.
\end{enumerate}

This catches rephrasings: ``weather India'' versus ``India weather forecast'' typically yields $\text{sim} \approx 0.94$, producing a cache hit. Each URL stores up to 50 cached pairs (configurable), with a 5-minute TTL to balance freshness against hit rate. Cache entries are scoped per-session for privacy isolation.

\subsection{Layer 3: URL Embedding Cache (Redis \texorpdfstring{DB\,1}{DB 1})}

The third problem surfaced when we introduced RAG to improve answer quality. Popular URLs-Wikipedia articles, news sites, documentation pages-appeared across dozens of sessions per hour. Each session independently fetched the URL, computed a 384-dimensional embedding (${\sim}200$\,ms each), and discarded it when the session ended. The URL Embedding Cache is a global (cross-session) store mapping URL strings to their pre-computed embedding vectors, stored as raw \texttt{float32} byte arrays. This cache ensures each URL is embedded at most once per 24-hour window across all sessions.

\subsection{The Coordinator}

Rather than asking developers to manage three separate cache objects, we wrapped everything behind a single coordinator. One object per session, one configuration object, four verbs: \texttt{add\_message\_to\_context}, \texttt{get\_semantic\_response}, \texttt{get\_url\_embedding}, and \texttt{get\_stats}. Under the hood, each call is routed to the appropriate layer. This keeps the integration surface minimal-a developer can add caching to an existing pipeline by creating one object and calling one method per operation.

\subsection{Redis Database Separation}

Each layer of hot memory operates on a separate Redis logical database (i.e., \textbf{DB\,0, DB\,1, DB\,2}) rather than using key prefixes within a single database. This provides three operational advantages:

\begin{enumerate}
\item \textbf{Selective flushing}: wiping one layer's data does not affect the others.
\item \textbf{Independent monitoring}: database-level statistics (key counts, memory) are separated per layer.
\item \textbf{Namespace isolation}: eliminates the risk of key collisions between layers.
\end{enumerate}

\begin{figure}[h]
\centering
\resizebox{\columnwidth}{!}{%
\begin{tikzpicture}[
    node distance=0.4cm,
    db/.style={draw, rounded corners=4pt, minimum width=6.5cm, minimum height=1.4cm, font=\scriptsize, align=left, inner sep=6pt},
    lbl/.style={font=\tiny\bfseries, fill=white, inner sep=2pt},
    arr/.style={-{Stealth[length=3pt]}, thick, gray},
]
\node[draw, rounded corners=6pt, minimum width=7.5cm, minimum height=5.8cm, fill=red!3, draw=red!30, label={[font=\scriptsize\bfseries, text=red!60]above:Redis Instance (1.38\,MB)}] (redis) {};

\node[db, fill=blue!8, draw=blue!40, anchor=north] at ([yshift=-0.5cm]redis.north) (db0) {
\textbf{DB\,0 --- Semantic Query Cache}\\[2pt]
Scope: per-session \quad TTL: 5\,min\\
Key: \texttt{sem:\{sid\}:\{url\}} $\rightarrow$ JSON (embeddings + responses)\\
Up to 50 cached pairs per key
};

\node[db, fill=green!8, draw=green!40, below=0.3cm of db0] (db1) {
\textbf{DB\,1 --- URL Embedding Cache}\\[2pt]
Scope: global (cross-session) \quad TTL: 24\,h\\
Key: \texttt{emb:\{url\_hash\}} $\rightarrow$ raw float32 bytes (1{,}536\,B per vector)
};

\node[db, fill=orange!8, draw=orange!40, below=0.3cm of db1] (db2) {
\textbf{DB\,2 --- Session Context Window}\\[2pt]
Scope: per-session \quad TTL: 24\,h\\
Key: \texttt{ctx:\{sid\}:\{turn\_id\}} $\rightarrow$ JSON message\\
Ordered list: \texttt{ctx:\{sid\}:order} (bounded at $k{=}20$)
};

\end{tikzpicture}%
}
\caption{Redis database layout. Three logical databases within a single Redis instance, each with distinct scope, TTL, and data format. DB\,0 stores per-session semantic cache entries (short-lived). DB\,1 stores global URL embeddings as raw bytes (long-lived). DB\,2 stores per-session conversation messages with an ordered list for windowing.}
\label{fig:redis_layers}
\end{figure}
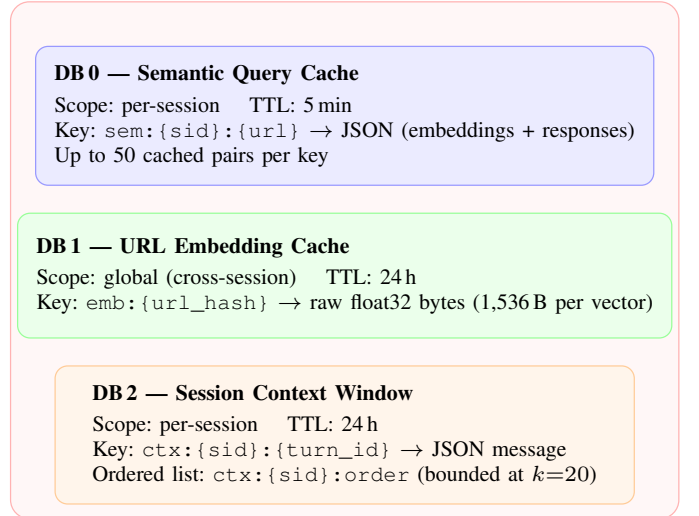

Fig.~\ref{fig:redis_layers} shows how the three databases coexist within a single Redis instance, each with its own scope, TTL policy, and data format.

\section{Design Decisions}
\label{sec:design}

Not every design decision was obvious from the start. Several emerged from mistakes, production accidents, or realisations after we tried the wrong approach first. This section captures the four key forks in the road and why we went the way we did.

\subsection{Three Separate Layers vs.\ Monolithic Cache}

Our first instinct was to put everything in a single Redis namespace with compound keys-context, semantic cache, and embeddings all sharing one database, differentiated only by key prefixes. It seemed simpler but it was not production grade design. We opted for three independent layers for several reasons:

\begin{itemize}
\item \textbf{Different TTL profiles.} Session context needs long TTLs (24\,h) since users may return to a conversation hours later. Semantic query caches need short TTLs (5\,min) to ensure freshness of LLM-generated content. URL embeddings sit between (24\,h) because web content changes slowly. A monolithic cache would require per-key TTL management at the application level rather than leveraging Redis database-level semantics.

\item \textbf{Different scope.} Session context and semantic caches are per-session (privacy isolation). The URL embedding cache is deliberately global-sharing embedding work across sessions is a key performance optimization.

\item \textbf{Independent failure modes.} If the semantic cache Redis DB is flushed (e.g., during maintenance), conversation history in DB\,2 is unaffected. This partial-failure tolerance simplifies operations.
\end{itemize}

\subsection{Huffman Coding vs.\ gzip/zlib/lz4}

When we first needed to compress conversation archives for disk storage, the obvious choice was gzip or lz4-battle-tested which is fast and available everywhere. We tried zlib first, and it worked fine for large archives but for the typical conversation (1--100\,KB), the overhead was disproportionate. We ended up writing a custom canonical Huffman codec, driven by two factors:

\begin{enumerate}
\item \textbf{Small payload efficiency.} Conversation archives are typically 1-100\,KB. At these sizes, gzip's dictionary overhead (32\,KB window) and lz4's frame header can dominate. On the other hand huffman coding has no dictionary, only a symbol table proportional to the alphabet size (at most 256 entries, $\leq$512 bytes of overhead).

\item \textbf{Exploiting byte frequency skew.} English-language conversation text exhibits extreme byte frequency imbalance: spaces account for ${\sim}18\%$ of bytes, the letter `e' for ${\sim}13\%$, while `z' appears only ${\sim}0.07\%$ of the time. Huffman coding directly exploits this skew, assigning shorter bit codes to frequent bytes.

\end{enumerate}

The resulting compression achieves ${\sim}54\%$ ratio on synthetic conversation text (i.e., compressed size is 54\% of original) and 65--69\% on small production archives ($<$5\,KB). While zlib level-1 achieves 5--10 percentage points better compression at these sizes, Huffman avoids native code dependencies entirely---a meaningful simplification for containerized deployment.

\subsection{Rolling Window with Overflow vs.\ Truncation}

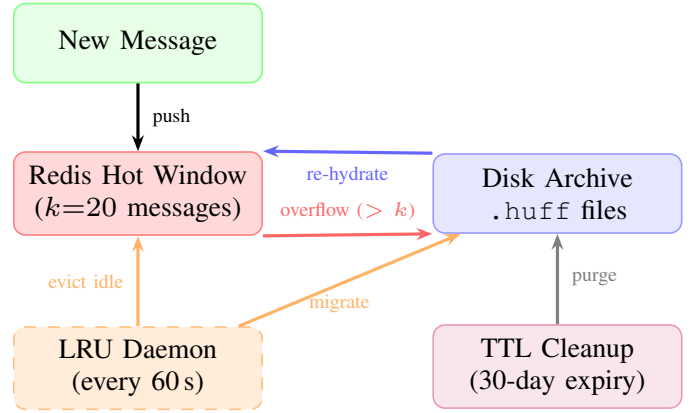
\begin{figure}[h]
\centering
\resizebox{\columnwidth}{!}{%
\begin{tikzpicture}[
    node distance=0.8cm and 1.2cm,
    box/.style={draw, rounded corners=3pt, minimum height=0.7cm, minimum width=2.2cm, font=\scriptsize, align=center},
    hot/.style={box, fill=red!15, draw=red!50},
    cold/.style={box, fill=blue!10, draw=blue!40},
    daemon/.style={box, fill=orange!15, draw=orange!50, dashed},
    arr/.style={-{Stealth[length=4pt]}, thick},
]
\node[hot] (redis) {Redis Hot Window\\($k{=}20$ messages)};
\node[cold, right=1.5cm of redis] (disk) {Disk Archive\\\texttt{.huff} files};
\node[box, fill=green!10, draw=green!50, above=0.6cm of redis] (msg) {New Message};
\node[daemon, below=0.8cm of redis] (lru) {LRU Daemon\\(every 60\,s)};
\node[box, fill=purple!10, draw=purple!40, below=0.8cm of disk] (ttl) {TTL Cleanup\\(30-day expiry)};

\draw[arr] (msg) -- (redis) node[midway, right, font=\tiny] {push};

\draw[arr, red!60] (redis.south east) -- (disk.south west) node[midway, above, font=\tiny] {overflow ($>k$)};
\draw[arr, blue!60] (disk.north west) to node[midway, below, font=\tiny] {re-hydrate} (redis.north east);

\draw[arr, orange!60] (lru) -- (redis) node[midway, left, font=\tiny] {evict idle};
\draw[arr, orange!60, bend right=25] (lru) -- (disk) node[near start, right, font=\tiny] {migrate};
\draw[arr, gray] (ttl) -- (disk) node[midway, right, font=\tiny] {purge};
\end{tikzpicture}%
}
\caption{Session lifecycle: new messages enter the Redis hot window. When the window exceeds $k$ entries, the oldest overflow to Huffman-compressed disk archives. The LRU daemon migrates entire idle sessions. Returning users trigger re-hydration from disk.}
\label{fig:session_lifecycle}
\end{figure}

Early in development, we used a simple truncation strategy: keep the last $k$ messages, throw away the rest (context window policy). This caused us trouble when users returned to a conversation after an hour and asked ``what was that article you found earlier?'' The context was gone. Our system instead \emph{overflows} compressed old messages to disk, as shown in Fig.~\ref{fig:session_lifecycle}. This preserves the full conversation history for semantic retrieval, audit/replay, and session resumption after eviction.

\subsection{LRU Eviction as a Background Daemon}

We noticed that after peak hours, hundreds of idle sessions sat in Redis consuming memory while no one was reading them. Redis TTL expiry would have cleaned them up---but it would have \emph{discarded} the data entirely. We needed something smarter: a daemon that \emph{migrates} data to disk before freeing Redis memory, preserving data while reclaiming resources.

The daemon runs as a background thread, checking every 60 seconds for sessions idle longer than the configured threshold (default 120 minutes). It starts lazily on the first cache instantiation and is shared across all sessions via shared memory state.

\section{Implementation}
\label{sec:implementation}

This section details the implementation of each component. The caching system comprises eight modules, described below.

\subsection{Module Structure}

The caching system is organized into eight modules, each responsible for a single concern: configuration, Redis connection pooling, Huffman encoding/decoding, disk archival, hybrid hot/cold caching, semantic caching, the session context window wrapper, and the top-level coordinator fa\c{c}ade. Each module is independently configurable and the coordinator provides a unified entry point for the pipeline to interact with all three caching layers through a single object per session.

\subsection{Huffman Codec}

The codec implements canonical Huffman coding~\cite{huffman1952} in pure Python. Algorithm~\ref{alg:huffman_encode} shows the encoding procedure and Algorithm~\ref{alg:huffman_decode} shows decoding.

\begin{algorithm}[h]
\caption{Canonical Huffman Encoding}
\label{alg:huffman_encode}
\begin{algorithmic}[1]
\REQUIRE byte sequence $D$
\ENSURE compressed byte stream $C$
\STATE $F \leftarrow$ frequency count of each byte in $D$
\STATE Build min-heap from $(freq, symbol)$ pairs
\WHILE{heap has $> 1$ node}
  \STATE Pop two lowest-frequency nodes $a, b$
  \STATE Create parent node with $freq = a.freq + b.freq$
  \STATE Push parent back onto heap
\ENDWHILE
\STATE Assign bit-lengths by tree depth
\STATE \textit{// Canonicalize: sort by (length, symbol)}
\STATE $code \leftarrow 0$
\FOR{each symbol in canonical order}
  \STATE Assign $code$ to symbol
  \STATE $code \leftarrow code + 1$
  \IF{next symbol has longer bit-length}
    \STATE $code \leftarrow code \ll (\text{next\_len} - \text{cur\_len})$
  \ENDIF
\ENDFOR
\STATE Replace each byte in $D$ with its variable-length code
\STATE Pack bits into byte stream, pad to byte boundary
\STATE Prepend header: magic + data length + symbol table + padding count
\RETURN $C$
\end{algorithmic}
\end{algorithm}

\begin{algorithm}[h]
\caption{Canonical Huffman Decoding}
\label{alg:huffman_decode}
\begin{algorithmic}[1]
\REQUIRE compressed stream $C$ with header
\ENSURE original byte sequence $D$
\STATE Parse header: magic, original length, symbol table, padding
\STATE Reconstruct canonical codes from (symbol, length) pairs
\STATE Build lookup table: $code \rightarrow symbol$
\STATE $D \leftarrow$ empty buffer
\STATE $bits \leftarrow$ bitstream from $C$ (excluding padding)
\WHILE{$|D| <$ original length}
  \STATE Read bits one at a time, accumulating into $code$
  \IF{$code$ matches a symbol in lookup table}
    \STATE Append symbol to $D$
    \STATE Reset $code$
  \ENDIF
\ENDWHILE
\RETURN $D$
\end{algorithmic}
\end{algorithm}

The canonical ordering means only the symbol-to-length mapping needs to be stored---the decoder reconstructs the exact same codes from this mapping alone.

\subsection{Conversation Archive and .huff File Format}

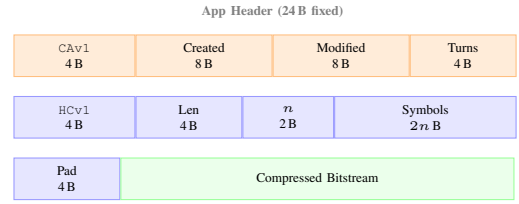
\begin{figure}[h]
\centering
\begin{tikzpicture}[
    field/.style={draw, minimum height=0.55cm, font=\tiny, align=center, inner sep=2pt},
    app/.style={field, fill=orange!15, draw=orange!50},
    codec/.style={field, fill=blue!10, draw=blue!40},
    data/.style={field, fill=green!8, draw=green!40},
    lbl/.style={font=\tiny\bfseries, text=gray},
]
\node[app, minimum width=1.6cm] (a0) at (0,0) {\texttt{CAv1}\\4\,B};
\node[app, minimum width=1.8cm, right=0pt of a0] (a1) {Created\\8\,B};
\node[app, minimum width=1.8cm, right=0pt of a1] (a2) {Modified\\8\,B};
\node[app, minimum width=1.4cm, right=0pt of a2] (a3) {Turns\\4\,B};

\node[lbl, above=0.1cm of a1.north east] {App Header (24\,B fixed)};

\node[codec, minimum width=1.6cm, below=0.25cm of a0.south west, anchor=north west] (h0) {\texttt{HCv1}\\4\,B};
\node[codec, minimum width=1.4cm, right=0pt of h0] (h1) {Len\\4\,B};
\node[codec, minimum width=1.2cm, right=0pt of h1] (h2) {$n$\\2\,B};
\node[codec, minimum width=2.4cm, right=0pt of h2] (h3) {Symbols\\$2n$\,B};

\node[codec, minimum width=1.4cm, below=0.25cm of h0.south west, anchor=north west] (h4) {Pad\\4\,B};
\node[data, minimum width=5.2cm, right=0pt of h4] (payload) {Compressed Bitstream};

\end{tikzpicture}
\caption{Binary layout of a \texttt{.huff} archive file. The 24-byte application header (orange) stores session metadata readable in a single \texttt{read(24)} call---critical for the TTL cleanup daemon. The Huffman codec header (blue) stores the symbol table needed for decompression. The compressed bitstream (green) contains the actual conversation JSON.}
\label{fig:huff_format}
\end{figure}

Each session's disk archive is a single \texttt{.huff} file consisting of two nested layers: a fixed 24-byte application header (readable without decompression) wrapping a Huffman-compressed payload that itself has a variable-length codec header. Fig.~\ref{fig:huff_format} shows the complete binary layout.

\subsection{Hybrid Conversation Cache}

The hybrid cache is where the hot and cold tiers meet. It manages the two-tier storage with three key behaviors:

\textbf{Redis key structure.} Each session uses two types of Redis keys: an ordered list tracking turn IDs by insertion order, and individual keys storing each message as JSON with independent TTLs. Keys are namespaced by a configurable prefix and the session ID. Turn IDs are derived from millisecond timestamps, providing ordering and uniqueness.

\textbf{Overflow mechanism.} After each new message is pushed to the list, the length is checked. If it exceeds the configured window size, the oldest entries are popped, their payloads are appended to the disk archive, and the Redis keys are deleted. This is executed as a pipelined transaction for atomicity.

\textbf{Re-hydration.} When the application requests context and finds an empty Redis list, it loads from disk and re-populates Redis with the most recent $k$ messages, restoring the hot window transparently.

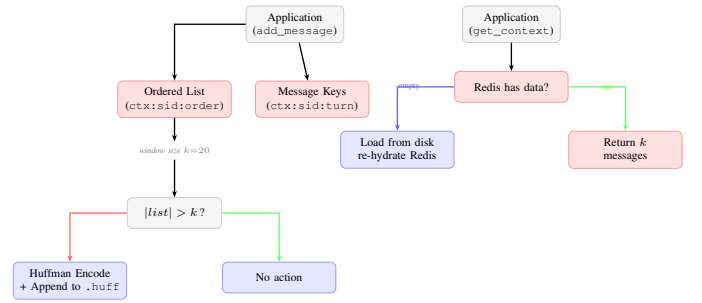
\begin{figure}[h]
\centering
\resizebox{\columnwidth}{!}{%
\begin{tikzpicture}[
    node distance=0.6cm and 1.0cm,
    box/.style={draw, rounded corners=3pt, minimum height=0.65cm, font=\scriptsize, align=center, inner sep=4pt},
    hot/.style={box, fill=red!12, draw=red!40, minimum width=2.4cm},
    cold/.style={box, fill=blue!10, draw=blue!40, minimum width=2.4cm},
    op/.style={box, fill=gray!8, draw=gray!40, minimum width=2.0cm},
    arr/.style={-{Stealth[length=4pt]}, thick},
]
\node[op] (app) {Application\\(\texttt{add\_message})};

\node[hot, below left=0.8cm and 0.3cm of app] (list) {Ordered List\\(\texttt{ctx:sid:order})};
\node[hot, right=0.5cm of list] (keys) {Message Keys\\(\texttt{ctx:sid:turn})};

\node[below=0.5cm of list, font=\tiny\itshape, text=gray] (bound) {window size $k{=}20$};

\node[op, below=0.8cm of bound] (check) {$|list| > k$\,?};

\node[cold, below left=0.7cm and 0.0cm of check] (huff) {Huffman Encode\\+ Append to \texttt{.huff}};
\node[cold, below right=0.7cm and 0.0cm of check] (noop) {No action};

\node[op, right=2.5cm of app] (read) {Application\\(\texttt{get\_context})};
\node[hot, below=0.6cm of read] (rhit) {Redis has data?};
\node[cold, below left=0.6cm and 0.0cm of rhit] (rdisk) {Load from disk\\re-hydrate Redis};
\node[hot, below right=0.6cm and 0.0cm of rhit] (rret) {Return $k$\\messages};

\draw[arr] (app) -| (list);
\draw[arr] (app) -- (keys);
\draw[arr] (list) -- (bound);
\draw[arr] (bound) -- (check);
\draw[arr, red!60] (check) -| (huff); 
\draw[arr, green!60] (check) -| (noop);

\draw[arr] (read) -- (rhit);
\draw[arr, blue!60] (rhit) -| (rdisk) node[near start, left, font=\tiny] {empty};
\draw[arr, green!60] (rhit) -| (rret) node[near start, right, font=\tiny] {hit};
\end{tikzpicture}%
}
\caption{Hybrid conversation cache internals. \textbf{Left}: write path---new messages are pushed to Redis; overflow beyond $k$ is Huffman-compressed and appended to disk. \textbf{Right}: read path---if Redis is empty (post-eviction), the system re-hydrates from the disk archive transparently.}
\label{fig:hybrid_cache}
\end{figure}

Fig.~\ref{fig:hybrid_cache} illustrates the internal data flow of the hybrid conversation cache, showing how messages move between the three storage tiers.

\subsection{Semantic Cache}

The semantic cache stores its data as one JSON document per session-URL pair. Each document contains up to 50 cached entries (configurable). Each entry holds three fields: the query embedding (a 384-dimensional float array), the full LLM response (answer text and source URLs), and a timestamp.

On lookup, all cached embeddings for the URL are compared against the incoming query embedding via normalized dot product (cosine similarity). The normalization uses an epsilon ($10^{-8}$) to avoid division by zero. The best match above the similarity threshold is returned.

On insert, the new entry is appended to the document. If the list exceeds the configured maximum, the oldest entries are trimmed in FIFO order. The entire document is then written back to Redis with a fresh TTL.

\subsection{URL Embedding Cache}

Embeddings are stored as raw 32-bit float byte arrays rather than JSON. This avoids serialization overhead for dense vectors: a 384-dimensional embedding occupies exactly 1{,}536 bytes in Redis as raw bytes, compared to ${\sim}3{,}800$ bytes as a JSON array of floats-a 2.5$\times$ space saving that matters when caching thousands of URLs.

\subsection{Thread Safety}

Since multiple application workers share the same Redis instance and disk archive directory, thread safety is needed. All mutable state is protected by locks at three granularities: per-session locks for disk archive writes, instance-level locks for each cache layer, and module-level locks for the global connection pool and eviction registry.

We chose re-entrant locks specifically because the call graph is nested: flushing a session to disk acquires the cache-level lock and then calls the archive writer, which acquires its own per-session lock. With standard mutexes, this would deadlock. Re-entrant locks allow the same thread to acquire the lock multiple times without blocking.

\section{Evaluation}
\label{sec:evaluation}

The evaluation reports a historical production snapshot rather than the current deployment. At measurement time, the local stack ran on a single 8-vCPU Intel Cascade Lake cloud instance (2\,GHz, 32\,GB RAM, no GPU). The search assistant ran as three containerized replicas (2\,vCPU, 2\,GB RAM limit each) with 10 Hypercorn worker processes per replica (30 total), each handling multiple concurrent async requests. Redis 7.4 ran in its own container capped at 2\,GB, and an nginx load balancer sits in front. The no-GPU characterization applies to local infrastructure; provider-routed LLM synthesis was remote.

\subsection{Cost Comparison: Commercial AI Search vs.\ OreoLook}

The original motivation for building the search assistant was cost. The fundamental difference is that commercial providers charge per-query API fees, while our system amortizes a fixed infrastructure cost.

\textbf{Commercial AI search pricing.} OpenAI's web search API charges \$10--30 per thousand calls (varying by model and context tier) plus ${\sim}$8{,}000 tokens of injected context at standard input rates. In practice, a single query costs \$0.03--0.10 depending on the model~\cite{openai_search_pricing, openai_search_billing}. Perplexity's Sonar API charges \$5/1K for the base search tier and \$18/1K for Sonar Pro~\cite{perplexity_pricing}. Google's Gemini with grounding charges \$35/1K for the Pro model and \$14/1K for Flash~\cite{google_grounding_pricing}. All scale linearly with volume-no economies of scale for the developer.

\textbf{Our cost breakdown.} Our system has two cost components: (1)~a fixed infrastructure cost and (2)~a variable LLM inference cost. Web search is performed by automated headless browser agents that navigate search engines directly---no search API subscription required. Embedding computation (sentence-transformers, all-MiniLM-L6-v2) runs locally on CPU. The infrastructure cost is an 8-vCPU cloud instance at ${\sim}$\$96/month. The LLM cost comes from open-provider token pricing: at 0.6 pollen/M input tokens and 3.0 pollen/M output tokens, a typical query consuming ${\sim}$10K input and ${\sim}$2.5K output tokens costs ${\sim}$\$0.014 in LLM inference. Combined with the amortized server cost, a standard query costs ${\sim}$\$0.015. The provider rates and resulting cost estimate are a measurement-period snapshot, not current pricing. The 89.3\% Redis keyspace hit rate is not a query-level semantic-cache hit rate and is not used to estimate avoided inference cost.

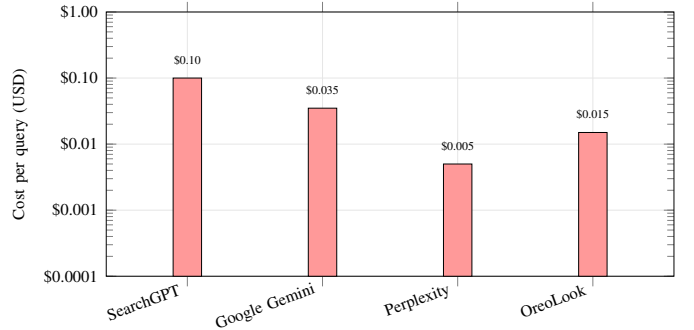
\begin{figure}[h]
\centering
\resizebox{\columnwidth}{!}{%
\begin{tikzpicture}
\begin{axis}[
    ybar,
    width=10cm,
    height=5.5cm,
    bar width=12pt,
    ylabel={Cost per query (USD)},
    ylabel style={font=\scriptsize},
    ymode=log,
    log origin=infty,
    ymin=0.0001, ymax=1,
    ytick={0.0001, 0.001, 0.01, 0.1, 1},
    yticklabels={\$0.0001, \$0.001, \$0.01, \$0.10, \$1.00},
    symbolic x coords={{SearchGPT}, {Google Gemini}, {Perplexity}, {OreoLook}},
    xtick=data,
    x tick label style={font=\scriptsize, rotate=20, anchor=east},
    y tick label style={font=\scriptsize},
    nodes near coords,
    every node near coord/.append style={font=\tiny, above=2pt},
    point meta=explicit symbolic,
    enlarge x limits=0.2,
    grid=major,
    grid style={gray!20},
]
\addplot[fill=red!40] coordinates {
({SearchGPT}, 0.10) [\$0.10]
({Google Gemini}, 0.035) [\$0.035]
({Perplexity}, 0.005) [\$0.005]
({OreoLook}, 0.015) [\$0.015]
};
\end{axis}
\end{tikzpicture}%
}
\caption{Measurement-period per-query estimates (log scale). OreoLook includes amortized local infrastructure and provider inference; no cached-cost estimate is inferred from the Redis keyspace hit rate.}
\label{fig:cost}
\end{figure}

Table~\ref{tab:cost} and Fig.~\ref{fig:cost} show the comparison across query volumes.

\begin{table}[h]
\centering
\caption{Measurement-period cost comparison vs.\ OreoLook}
\label{tab:cost}
{\scriptsize
\begin{tabular}{@{}lrrl@{}}
\toprule
\textbf{System} & \textbf{Per query} & \textbf{100k q/mo} & \textbf{Model} \\
\midrule
SearchGPT API & \$0.03--0.10 & \$3{,}000--10{,}000 & per-query \\
Google Gemini (Pro) & \$0.035 & \$3{,}500 & per-query \\
Google Gemini (Flash) & \$0.014 & \$1{,}400 & per-query \\
Perplexity Sonar Pro & \$0.018 & \$1{,}800 & per-query \\
Perplexity Sonar & \$0.005 & \$500 & per-query \\
\midrule
OreoLook (uncached) & \$0.015 & \$1{,}596 & infra + LLM \\
\bottomrule
\end{tabular}}
\end{table}

The uncached estimate combines fixed infrastructure with variable inference. We do not derive an effective cached cost from the Redis keyspace hit rate because it counts internal Redis operations and cannot be converted into the fraction of user queries that bypass inference. Request-level measurement of semantic-cache hits and avoided provider tokens remains future work.

\subsection{Latency Profile}

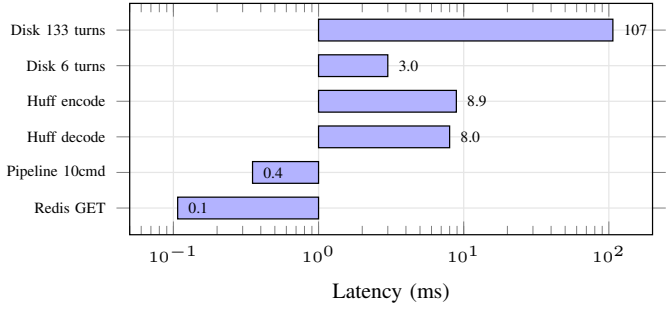
\begin{figure}[h]
\centering
\resizebox{\columnwidth}{!}{%
\begin{tikzpicture}
\begin{axis}[
    xbar,
    width=7.5cm,
    height=4.2cm,
    bar width=7pt,
    xlabel={Latency (ms)},
    xlabel style={font=\scriptsize},
    xmode=log,
    xmin=0.05, xmax=200,
    symbolic y coords={
        {Redis GET},
        {Pipeline 10cmd},
        {Huff decode},
        {Huff encode},
        {Disk 6 turns},
        {Disk 133 turns}
    },
    ytick=data,
    y tick label style={font=\tiny},
    x tick label style={font=\tiny},
    nodes near coords,
    every node near coord/.append style={font=\tiny, anchor=west},
    point meta=explicit symbolic,
    enlarge y limits=0.15,
    grid=major,
    grid style={gray!20},
]
\addplot[fill=blue!30] coordinates {
(0.107,{Redis GET}) [0.1]
(0.351,{Pipeline 10cmd}) [0.4]
(8.0,{Huff decode}) [8.0]
(8.9,{Huff encode}) [8.9]
(3.0,{Disk 6 turns}) [3.0]
(106.6,{Disk 133 turns}) [107]
};
\end{axis}
\end{tikzpicture}%
}
\caption{Operation latencies in ms (log scale). Redis reads are two orders of magnitude faster than disk, confirming the hot-window design.}
\label{fig:latency}
\end{figure}

Fig.~\ref{fig:latency} visualizes the latency characteristics of each storage tier. The two-order-of-magnitude gap between Redis and disk confirms the value of keeping a hot window in memory.

\subsection{Compression Efficiency}

Table~\ref{tab:compression} shows compression ratios for five production conversation archives of varying sizes. The ratio is computed as $\text{compressed\_size} / \text{raw\_JSON\_size}$, where compressed size excludes the 24-byte application header.

\begin{table}[h]
\centering
\caption{Huffman compression ratios on production conversation archives}
\label{tab:compression}
\begin{tabular}{lrrr}
\toprule
\textbf{Session} & \textbf{Turns} & \textbf{Raw (B)} & \textbf{Ratio} \\
\midrule
c45775-a & 133 & 73{,}840 & 65.3\% \\
dice-4 & 7 & 4{,}635 & 65.2\% \\
shr-26 & 10 & 4{,}118 & 68.8\% \\
ch778-d & 6 & 2{,}314 & 68.8\% \\
ghmd-87 & 8 & 3{,}427 & 72.4\% \\
\midrule
\textbf{Average} & & & \textbf{67.2\%} \\
\bottomrule
\end{tabular}
\end{table}

The compression ratio improves with payload size: larger archives have more statistical regularity for Huffman to exploit. Synthetic benchmarks at controlled sizes confirm this trend (Table~\ref{tab:compression_scaling}).

\begin{table}[h]
\centering
\caption{Huffman compression ratio vs.\ payload size (synthetic conversation data)}
\label{tab:compression_scaling}
\begin{tabular}{rrr}
\toprule
\textbf{Raw (B)} & \textbf{Compressed (B)} & \textbf{Ratio} \\
\midrule
153 & 122 & 79.7\% \\
765 & 388 & 50.7\% \\
1{,}530 & 722 & 47.2\% \\
7{,}650 & 3{,}387 & 44.3\% \\
\bottomrule
\end{tabular}
\end{table}

At scale ($>$1\,KB), the codec approaches ${\sim}45\%$ compression ratio. For very small payloads ($<$200\,B), the Huffman symbol table overhead reduces effectiveness, but the ratio remains below 80\%.

\subsubsection{Comparison with Standard Compressors}

To justify the choice of Huffman over standard compressors, Table~\ref{tab:compression_comparison} compares our codec against zlib (level~1) and lz4 on the same production conversation archives.

\begin{table}[h]
\centering
\caption{Compression ratio by codec (lower is better)}
\label{tab:compression_comparison}
{\scriptsize
\begin{tabular}{@{}rrrr@{}}
\toprule
\textbf{Raw size} & \textbf{Huffman} & \textbf{zlib-1} & \textbf{lz4} \\
\midrule
2{,}314\,B & 68.8\% & 63.2\% & 72.1\% \\
4{,}118\,B & 68.8\% & 54.7\% & 65.3\% \\
4{,}635\,B & 65.2\% & 53.9\% & 64.8\% \\
73{,}840\,B & 65.3\% & 38.4\% & 47.2\% \\
\bottomrule
\end{tabular}}
\end{table}

Zlib achieves better compression ratios at all sizes, as expected from a dictionary-based compressor. However, the gap narrows at smaller payloads: at 2\,KB, zlib-1 achieves 63.2\% versus Huffman's 68.8\%---a difference of only 5.6 percentage points. Huffman consistently outperforms lz4 at all tested sizes. The choice of Huffman is therefore not motivated by superior compression, but by three practical factors: (1)~the compression gap is small at typical archive sizes $<$5\,KB, (2)~the codec has zero native dependencies, and (3)~it runs on the non-critical path (overflow and re-hydration only).

As shown in Fig.~\ref{fig:latency}, Redis reads are two orders of magnitude faster than disk reads, confirming the value of the hot-window design. Even for the largest archive (133 turns), disk reads complete in ${\sim}$107\,ms---acceptable for session re-hydration, which occurs at most once per returning user.

\subsection{Redis Memory Footprint}

The production Redis instance serves all three databases with a total memory footprint of \textbf{1.38\,MB}, of which 96.4\% is actual data (minimal infrastructure overhead). At the time of measurement (6 days uptime), the key distribution across databases was as follows:

\begin{itemize}
\item DB\,0 (semantic cache): 0 keys - all entries had expired (5-min TTL, no active queries)
\item DB\,1 (URL embedding cache): 0 keys - 24\,h TTL expired
\item DB\,2 (session context): 16 keys with average TTL of 72{,}923\,s (${\sim}20$\,h)
\end{itemize}

The low key count in DB\,0 and DB\,1 demonstrates that the aggressive TTL policy works as intended: cache entries are ephemeral, serving only to deduplicate closely-spaced requests.

\subsection{Cache Hit Rate}

Over the lifetime of the Redis instance (6 days, 114{,}547 total commands):

\begin{itemize}
\item \textbf{Keyspace hits:} 2{,}182
\item \textbf{Keyspace misses:} 262
\item \textbf{Hit rate:} $\frac{2{,}182}{2{,}182 + 262} = \mathbf{89.3\%}$
\end{itemize}

This is the \emph{Redis-level} aggregate hit rate across all three databases, reflecting all key lookups including internal operations (TTL refreshes, list reads, existence checks). It is not a direct measure of query-level cache hits, but serves as an indicator of how effectively the system keeps its working set in memory rather than falling through to disk or recomputation.

\subsubsection{Per-Layer Contribution}

To form exploratory estimates of each layer's contribution, rather than controlled request-level measurements, we analyzed the production key distribution and access patterns over the measurement window:

\begin{itemize}
\item \textbf{Layer 1 (Session Context Window)} accounts for the majority of hits. Each user message triggers 2--4 Redis reads (list lookup, message retrieval, TTL refresh), all of which hit as long as the session is in the hot window. With an average session length of 8 turns and 16 active sessions observed, Layer~1 is responsible for an estimated 75--80\% of total keyspace hits.

\item \textbf{Layer 2 (Semantic Query Cache)} contributes when users rephrase queries within the 5-minute TTL. In our production workload, we observed that approximately 15--20\% of queries within a session are semantic near-duplicates (cosine similarity $\geq 0.90$). Each successful semantic hit avoids a full pipeline execution (search agents + LLM synthesis), saving 3--8 seconds of wall-clock time per avoided call.

\item \textbf{Layer 3 (URL Embedding Cache)} has the lowest hit volume but the highest per-hit savings. Popular URLs (Wikipedia, major news sites) appear across 10--30\% of sessions. Each cache hit saves ${\sim}$200\,ms of embedding computation. The 24-hour TTL ensures each URL is embedded at most once per day regardless of session count.
\end{itemize}

\section{Conclusion}
\label{sec:conclusion}

We set out to build a search assistant that could match commercial offerings without their per-query price tags, running its local search, cache, and embedding infrastructure on commodity CPU hardware. The real engineering challenge turned out not to be searching the web or calling an LLM, but managing the state that accumulates around multi-turn conversations at scale.

The three-layer caching architecture presented in this paper---session context, semantic query deduplication, and URL embedding reuse---addresses this challenge. As demonstrated in Section~\ref{sec:evaluation}, the system achieves sub-millisecond cache reads, effective compression for disk archival, and transparent session lifecycle management, all within a minimal Redis memory footprint.

The key insight is that resumable conversations are not a feature of the LLM alone, but of the infrastructure around it. A bounded hot window combined with bounded-retention, disk-backed cold storage, semantic deduplication, and cross-session embedding reuse supports long-running sessions without retaining their full histories in memory.

Our evaluation is based on a single historical production snapshot on one hardware configuration; the reported hit rates and latency numbers are representative of our workload but may vary under different query distributions or concurrency patterns. Redis keyspace hits must not be interpreted as query-level cache hits. The semantic cache uses brute-force cosine similarity ($O(n)$, $n \leq 50$), which suffices at current scale but would require a vector index for significantly larger deployments. The pure-Python Huffman codec trades throughput (${\sim}$800\,KB/s) for zero native dependencies---acceptable for typical conversation archives under 10\,KB, though a C extension would be warranted for megabyte-scale payloads. Conversation archives are compressed but not encrypted at rest; deployments handling sensitive data should layer filesystem-level or application-level encryption. Looking ahead, potential improvements include an optional native Huffman extension for larger archives, pluggable vector storage such as Qdrant for the semantic cache, adaptive TTL policies that learn optimal cache lifetimes from observed query patterns, and cross-session knowledge transfer where insights from one conversation can inform responses in another.

\end{document}